\documentstyle[prd,aps,12pt]{revtex}

\begin{document}
\draft
\newcommand{\Tr}{\mbox{Tr}}
\newcommand{\lTr}{\mbox{tr}}
\vspace{1.0cm}
%
\title{Screening length, dispersion relations and quark potential    
in thermo field dynamics}
%
%
%
\author{Wanyun Zhao \footnote{Permanent address: Institute of Theoretical 
Physics, Academia Sinica, Beijing, China.} 
 and F.C. Khanna 
\footnote{e-mail: khanna@phys.ualberta.ca}}
\address{Theoretical Physics Institute, University of Alberta, Edmonton, 
Alberta, Canada T6G 2J1 \\ 
 and \\
TRIUMF, 4004 Wesbrook Mall, Vancouver, British Columbia, Canada V6T 2A3}
%
%

\maketitle
%

\vspace{3cm}

\begin{abstract}
   The screening length in a quark-gluon plasma, the dispersion relations of
thermal gluon self-energy and the quark potential at high temperature are studied 
within thermo field dynamics framework. By calculation of the real and imaginary parts, 
of the gluon self-energy in one-loop order in thermo field dynamics, we obtain an 
expression for the screening length in a quark-gluon plasma and the dispersion relation 
between the real and imaginary parts. At high temperature, using photon exchange 
between electron-positron in a skeleton expansion and ladder approximation, the screened 
Coulomb potential is obtained, and using one-gluon and two-gluon exchange between a 
quark-antiquark, we get an expression for the screened quark potential up to $ O(g^{4}) $.
\end{abstract}

\widetext
\begin{center}

\section{INTRODUCTION}
\end{center}
 
  The CERN heavy-ion program and the forthcoming start of the BNL-RHIC
( Relativistic Heavy Ion Collider ) make the quark-gluon plasma (QGP) phase of     
hot QCD of great current interest [1]. The screening length and the potential 
between quarks in a QGP, and the dispersion relations of the self-energy of gluon 
are important topics for the study of the QGP phase in a hot QCD. Quantum field 
theory at finite temperature and density is a powerful tool for the study of 
these subjects. There are three different approaches to Quantum field theory at
finite temperature and density. It is mainly the following: the older Euclidean 
imaginary-time Matsubara formalism [2], the closed time-path method due to 
Schwinger and Keldysh [3], and the real time operator-based finite temperature
field theory, thermo field dynamics (TFD) due to Takahashi and Umezawa [4]. The 
central idea of TFD is the thermal vacuum and the doubling of the Hilbert space of 
states. It permits thermal averages, which are traces over Fock space, to be written
as expectation values with respect to the thermal vacuum. The operators in this 
doubled space are effectively doubled in number as well. Within TFD, the quantum 
field theory is naturally extended to finite temperature without having to consider 
imaginary time. TFD is more concise and an easy and convenient way to study the 
above subjects.

  The studies reported in this paper are based on TFD. Its purpose is three-fold: 
(1) within this framework we calculate perturbatively the screening length in the QGP 
at one-loop level; (2) the dispersion relations of self energy of gluon are 
derived and, (3) the potential between a quark and an antiquark in a hot QCD is obtained 
up to $ O (g^{4})$, $g$ being the QCD coupling constant. As a simple example the 
potential between electron and positron in a hot QED is derived.

  The paper is organized as follows: in Sec. II, the screening length in a hot QCD
is obtained in one-loop level in TFD approach, and we discuss both the 
color-electric and the color-magnetic sectors. In Sec. III, the dispersion 
relations of the gluon self energy at finite temperature are proved and explain the
physical results on the real part and imaginary part of the gluon self energy. In 
Sec. IV, we calculate the screened Coulomb potential in a hot QED up to all 
orders and the potential between a quark and an antiquark in a hot QCD up to 
$ O (g^{4}) $. Finally, in Sec. V some conclusions and comments are presented. \\    

\vspace{1cm}

\section{SCREENING LENGTH}

  In this section we study the gluon self-energy in one-loop order. To 
$ O (g^{2}) $, the gluon self-energy is given by the one-loop diagrams in Fig.1.

  The full gluon propagator $D^{ab}_{\mu\nu}(x-x') = < A^{a}_{\mu}(x) A^{b}_{\nu}(x') >$
may be expressed in terms of the one-particle irreducible self-energy,
$\Pi^{ab}_{\mu\nu}(k)$, 

\begin{equation}
D_{\mu\nu}^{ab}(k) = \frac{1}{k^{2}g_{\mu\nu}\delta^{ab} + 
\Pi_{\mu\nu}^{ab}(k)},
\end{equation}
where in Minkowski space $g_{\mu\nu}$ is the metric tensor with components

\begin{equation}
g_{00}=-g_{11}=-g_{22}=-g_{33}=1;  g_{\mu\nu}=0, \mu\not=\nu.
\end{equation}

  Note that $D^{ab}_{\mu\nu}$ is diagonal in the color index $a$ and $b$. 
$\Pi^{ab}_{\mu\nu}(k)$ is expressed as

\begin{equation}
\Pi^{ab}_{\mu\nu}(k) = \delta^{ab}(g_{\mu\nu}k^{2} - k_{\mu}k_{\nu})\Pi(k^{2}).
\end{equation}
where $\Pi^{ab}_{\mu\nu}(k)$ is transverse to one-loop order.




  Before we study the gluon self-energy in one-loop order within TFD framework, it is 
convenient to rewrite the standard propagators in TFD, $\Delta(k)$ for bosons and $S(p)$ for 
fermions, as follows [5]:

for gluons

\begin{eqnarray}
\Delta(k) & = & U_{B}(k_{0},\beta)\bar\Delta(k)U^{\dagger}_{B}(k_{0},\beta)
	\nonumber\\ 
          & = & \left( \begin{array}{cc}
\frac{g_{\mu\nu}}{k^{2}+i\epsilon}-\frac{(1-\lambda^{-1})k_{\mu}k_{\nu}}{(k^{2}+i\epsilon)^{2}} & 0\\
0 & -\frac{g_{\mu\nu}}{k^{2}-i\epsilon}+\frac{(1-\lambda^{-1})k_{\mu}k_{\nu}}{(k^{2}-i\epsilon)^{2}}
\end{array}
\right)-2\pi i\delta(k^{2})\frac{g_{\mu\nu}}{e^{\beta |k_{0}|}-1}\left(
\begin{array}{cc}
1 & e^{\frac{\beta |k_{0}|}{2}}\\
e^{\frac{\beta |k_{0}|}{2}} & 1
\end{array}\right)
\end{eqnarray}
where

\begin{equation}
\begin{array}{c}

\bar\Delta(k)=\frac{\tau}{(k_{0}+i\delta\tau)^{2}-\vec{k}^{2}}[g_{\mu\nu}-
\frac{(1-\lambda^{-1})k_{\mu}k_{\nu}}{(k_{0}+i\delta\tau)^{2}-\vec{k}^{2}}],
	\nonumber\vspace{0.3in} \\

        \tau =\left(
\begin{array}{cc}
1 & 0\\
0 & -1
\end{array}\right),
	\nonumber\vspace{0.3in}\\

U_{B}(k_{0},\beta) = \left(
\begin{array}{cc}
\mbox{cosh}\theta(k_{0},\beta)  &\mbox{sinh}\theta(k_{0},\beta)\\
\mbox{sinh}\theta(k_{0},\beta) & \mbox{cosh}\theta(k_{0},\beta)
\end{array}\right),    
	\nonumber\vspace{0.3in}\\

\mbox{sinh}^{2}\theta(k_{0},\beta)=\frac{1}{e^{\beta |k_{0}|}-1},

\end{array}
\end{equation}
where $\beta = \frac{1}{k_{B}T}$, $k_{B}$ and $T$ are the Boltzmann constant
and the temperature of the quark-gluon system respectively;

and for quarks

\begin{eqnarray}
S(p)&=&U_{F}(\vec{p},\beta)\bar S(p)U_{F}^{\dagger}(\vec{p},\beta)
             \nonumber\\
&=&(\not p+m)\left [\left (
\begin{array}{cc}
\frac{1}{p^{2}-m^{2}+i\epsilon} & 0\\
0 & \frac{1}{p^{2}-m^{2}-i\epsilon}
\end{array}\right)
+2\pi i\delta(p^{2}-m^{2})\frac{1}{e^{\beta |p_{0}|}+1}\left(
\begin{array}{cc}
1 & -\epsilon(p_{0})e^{\frac{\beta |p_{0}|}{2}}\\
-\epsilon(p_{0})e^{\frac{\beta |p_{0}|}{2}} & -1  
\end{array}\right)\right]
\end{eqnarray}
where

\begin{equation}
\begin{array}{c}
\bar S(p)=\frac{I}{\not p-m+i\delta\tau},\hspace{0.2in}

I=\left(
\begin{array}{cc}
1 & 0\\
0 & 1
\end{array}\right),\vspace{0.3in}\\
U_{F}(\vec p,\beta)=\left(
\begin{array}{cc}
\cos\theta(|\vec p|,\beta) & \epsilon(p_{0})\sin\theta(|\vec p|,\beta)\\
-\epsilon(p_{0})\sin\theta(|\vec p|,\beta) & \cos\theta(|\vec p|,\beta)
\end{array}\right),\hspace{0.2in}
\sin^{2}\theta(|\vec p|,\beta)=\frac{1}{e^{\beta p_{0}}+1},\vspace{0.3in}\\

and \hspace{0.2in}\epsilon(p_{0})=\theta(p_{0})-\theta(-p_{0}),\hspace{0.2in}
\theta(p_{0})=
\left\{
\begin{array}{c}
1, p_{0}>0,\\
0, p_{0}<0.
\end{array}
\right.

\end{array}
\end{equation}
The matrices $U_{B}$ and $U_{F}$ are the Bogoliubov 
transformation matrices [6] for bosons and fermions respectively .\\

  We calculate the contributions of diagrams in Fig.1 to the gluon 
self-energy in the Feynman gauge, $\lambda =1$, no ghost state appears.
In TFD, the operator of the gluon self-energy is a $(2\times 2)$ matrix in
the Hilbert space of states. Now we express explicitly the $(1,1)$ component 
of the contributions of the diagram $1(a)$ to the gluon self-energy.\\

\begin{eqnarray}
[\Pi_{\mu\nu}^{ab}(k)]_{Fig.1(a)}^{11} & = & g^{2}Q\delta^{ab}N_{f}\int\frac{d^{4}p}{(2\pi)^{4}}
tr\lbrace\gamma_{\mu}[(\not p-\not k)+m][\frac{1}{(p-k)^{2}-m^{2}+i\epsilon}
+2\pi i\delta[(p-k)^{2}-m^{2}] \nonumber \\
& & \frac{1}{e^{\beta |p_{0}-k_{0}|}+1}]\gamma_{\nu}(\not p +m)
[\frac{1}{p^{2}-m^{2}+i\epsilon}+2\pi i\delta(p^{2}-m^{2})\frac{1}{e^{\beta |p_{0}|}+1}]\rbrace \nonumber \\
& & = g^{2}Q\delta^{ab}N_{f}\int\frac{d^{4}p}{(2\pi)^{4}}tr\lbrace\gamma_{\mu}\frac{[(\not p-\not k)+m]}
{[(p-k)^{2}-m^{2}+i\epsilon]}\gamma_{\nu}\frac{(\not p +m)}{(p^{2}-m^{2}+i\epsilon)}\rbrace \nonumber \\
& & + 2g^{2}Q\delta^{ab}N_{f}\int\frac{d^{4}p}{(2\pi)^{4}}2\pi i\delta(p^{2}-m^{2})
tr\lbrace\gamma_{\mu}\frac{[(\not p-\not k)+m]}{[(p-k)^{2}-m^{2}+i\epsilon]}\gamma_{\nu}
\frac{(\not p +m)}{e^{\beta |p_{0}|}+1}\rbrace \nonumber \\
& & + g^{2}Q\delta^{ab}N_{f}\int\frac{d^{4}p}{(2\pi)^{4}}(-4\pi^{2})\delta[(p-k)^{2}-m^{2}]
\delta(p^{2}-m^{2}) \nonumber \\
& & tr\lbrace\gamma_{\mu}\frac{[(\not p-\not k)+m]}{e^{\beta |p_{0}-k_{0}|}+1}
\gamma_{\nu}\frac{(\not p +m)}{e^{\beta |p_{0}|}+1}\rbrace ,
\end{eqnarray}
where $Tr(T^{a}T^{b})=Q\delta^{ab}$, $T^{a}$, $T^{b}$ are generators of the non-abelian
gauge group, $N_{f}$ is the quark flavor number, and $Q$ is the Casimir operator related 
to the fermion representation of the non-abelian gauge group. We normalize the generators
$T^{a}$ and $T^{b}$, so that $Q=1$. The first term on the right side of Eq. (8) is 
independent of temperature, $T=0$. The second and third terms in the right side of Eq. (8)
depend on temperature, $T\not=0$. The second term is the real part of the gluon self-energy 
and the third term is the imaginary part of the gluon self-energy. In order to obtain the 
pole mass of a gluon at finite temperature, we consider the $(0,0)$ component of 
$\Pi_{\mu\nu}^{ab}$ and let the momentum $\vec{k}\rightarrow 0$ and the energy $k_{0}\rightarrow 0$.
For simplicity, we omit the quark mass $m$ in Eq. (8). Then, we get the contribution of 
the diagram $(a)$ in Fig.1 to the square of the pole mass of a gluon at finite temperature
from Eq. (8)

\begin{equation}
Re[\Pi_{00}^{ab}(\vec{k}\rightarrow 0, k_{0}\rightarrow 0, T\not=0)]_{Fig.1(a)}^{11}
=\frac{2g^{2}Q\delta^{ab}N_{f}}{\pi^{2}}\int_{0}^{\infty}\frac{d\omega\omega}{(e^{\beta\omega}
+1)},
\end{equation}
where in the integral for $p_{0}$, we have used this formula

\begin{equation}
\delta(p^{2}-m^{2})=\frac{1}{2\omega}[\delta(p_{0}+\omega)+\delta(p_{0}-\omega)],\hspace{0.2in} 
\omega=\sqrt{{\vec{p}}^{2}+m^{2}}.
\end{equation}
After using the following integral 

\begin{equation}
\int_{0}^{\infty}\frac{x^{\nu -1}dx}{e^{\mu x}+1}=\frac{1}{\mu^{\nu}}(1-2^{1-\nu})
\Gamma(\nu)\zeta(\nu), \hspace{0.2in}( Re\mu>0, Re\nu>0 )
\end{equation}
where $\Gamma(\nu)$ and $\zeta(\nu)$ are the Gamma function and the Riemann's zeta
function respectively, we get

\begin{equation}
Re[\Pi_{00}^{ab}(\vec{k}\rightarrow 0, k_{0}\rightarrow 0, T\not=0)]_{Fig.1(a)}^{11}
=\frac{2g^{2}Q\delta^{ab}N_{f}}{\pi^{2}}\frac{1}{\beta^{2}}\frac{1}{2}
\Gamma(2)\zeta(2)
\end{equation}

  Similarly we have obtained contributions of diagrams (b) and (d) in Fig.1 to the
square of the pole mass of a gluon at finite temperature, as follows:

\begin{equation}
Re[\Pi_{00}^{ab}(\vec{k}\rightarrow{0}, k_{0}\rightarrow{0}, T\not=0)]_{Fig.1(b),(d)}^{11}
=\frac{2g^{2}C\delta^{ab}}{\pi^{2}}\frac{1}{\beta^{2}}\Gamma(2)\zeta(2)
\end{equation}
where 

\begin{equation}
\sum_{a,b}C_{abc}C_{abd}=C\delta^{cd},\hspace{0.2in} C_{abc}=-C_{bac};\hspace{0.2in}
 C=N,\hspace{0.2in} for\hspace{0.1in} SU(N)\hspace{0.1in} gauge \hspace{0.05in} group.
\end{equation}
Here $C_{abc}$ are structure constants of the non-abelian gauge group, and $C$
is the Casimir operator related to the adjoint representation of the non-abelian gauge group.\\
In the calculation of Eq. (13) we have used the following formula:

\begin{equation}
\int_{0}^{\infty}\frac{x^{\nu -1}dx}{e^{\mu x}-1}=\frac{1}{\mu^{\nu}}\Gamma(\nu)\zeta(\nu)
\hspace{0.2in}( Re\mu>0, Re\nu>1 ).
\end{equation}

  To sum up, we obtain finally the total contributions of all diagrams in Fig.1
to the square of the pole mass of a gluon at finite temperature in one-loop order   

\begin{eqnarray}
Re[\Pi_{00}^{ab}(\vec{k}\rightarrow 0,k_{0}\rightarrow 0,T\not=0]_{Fig.1}^{11}
 & = & \delta^{ab}m_{el}^{2} \nonumber \\
 & = & \frac{2g^{2}T^{2}\delta^{ab}}{\pi^{2}}(N+\frac{1}{2}N_{f})
\Gamma(2)\zeta(2) \nonumber \\
 & = & \delta^{ab}\frac{1}{3}(N+\frac{1}{2})g^{2}T^{2},
\end{eqnarray}
where
\begin{equation}
\Gamma(2)=1, \hspace{0.3in} \zeta(2)=\sum_{n=1}^{\infty}\frac{1}{n^{2}}=\frac{\pi^{2}}{6},
\end{equation}
and the mass $m_{el}$ is the color electric mass of a gluon at finite temperature 
in one-loop order.

  By calculations for the spatial components of the gluon self-energy in Fig.1
in TFD, it is found that $\Pi_{i0}^{ab}(k_{0},\vec{k}=0,T\not=0), \Pi_{0i}^{ab}$
and $\sum_{i=1}^{3}\Pi_{ii}^{ab}$ are all zero. So, the color magnetic mass of a 
gluon at finite temperature in one-loop order equals zero. Therefore in one-loop 
order the color-electric gluon fields are screened and the color-magnetic gluon
fields are not screened. The conclusions from the present calculation of the gluon
self-energy in one-loop order in TFD are the same as in the conventional imaginary-time 
formalism (ITF) [7], but the calculations in TFD are more clear, simple and easy. 
The screening length in a QGP is the reciprocal of $m_{el}$, then

\begin{equation}
L(T)=\frac{\sqrt{3}}{\sqrt{N+\frac{1}{2}N_{f}}gT}.
\end{equation}\\


\section{DISPERSION RELATION}

  In this section we discuss what is the dispersion relation for the thermal
gluon self-energy? For this purpose, we consider the renormalized gluon thermal propagator\\

\begin{equation}
[\Delta_{R}^{-1}(k)]_{\mu\nu}^{\alpha\beta}=g_{\mu\nu}k^{2}\tau^{\alpha\beta}-\Pi_{\mu\nu}^{\alpha\beta}(k)
\end{equation}

\begin{equation}
[\Delta_{R}(k)]_{\mu\nu}^{\alpha\beta}=[U_{B}(k_{0})\frac{1}
{\bar\Delta_{\mu\nu}^{-1}(k)-\bar\Pi_{\mu\nu}(k)}U_{B}^{\dagger}(k_{0})]^{\alpha\beta}
\end{equation}
where the Feynman gauge has been chosen, $\lambda=1$, and $\Pi(k)$ and $\bar\Pi(k)$ 
are connected with each other through the following relation\\

\begin{equation}
\Pi^{\alpha\beta}(k)=[\tau U_{B}(k_{0})\bar\Pi(k)U_{B}(k_{0})\tau]^{\alpha\beta}
\end{equation}
where the $(\mu,\nu)$ vector indices are omitted and \\

\begin{equation}
\bar\Pi^{\alpha\beta}=\tau^{\alpha\beta}Re\bar\Pi+i\delta^{\alpha\beta}Im\bar\Pi.
\end{equation}
Explicitly, the $(1,1)$ component of the relation $(20)$ is \\

\begin{equation}
[\Delta_{R}(k)]_{\mu\nu}^{11}=g_{\mu\nu}\lbrack\frac{\mbox{cosh}^{2}\theta(k_{0})}
{(k_{0}+i\Gamma)^{2}-\omega^{2}}-\frac{\mbox{sinh}^{2}\theta(k_{0})}
{(k_{0}-i\Gamma)^{2}-\omega^{2}}\rbrack
\end{equation}

\begin{equation}
\omega^{2}=\vec{k}^{2}+m^{2}+Re\bar\Pi-\Gamma^{2}, \hspace{0.2in} Im\bar\Pi=-2k_{0}\Gamma
\end{equation}
From Eqs. (22) and (24) it is clear that the real and imaginary part 
of our interest are given by $Re\bar\Pi$ and $Im\bar\Pi$. From Eqs. (21) and
(22) we obtain \\

\begin{equation}
Re\Pi^{11}=-Re\Pi^{22}, \hspace{0.2in} Re\Pi^{12}=Re\Pi^{21}=0,
\end{equation}

\begin{equation}
Im\Pi^{11}=Im\Pi^{22}, \hspace{0.2in} Im\Pi^{12}=Im\Pi^{21},
\end{equation}

\begin{equation}
Im\Pi^{11}=-\frac{1}{2}\frac{e^{\beta |k_{0}|}+1}{e^{\beta |k_{0}|/2}}Im\Pi^{12}. 
\end{equation}
Further, we get \\

\begin{equation}   
Re\bar\Pi=Re\Pi^{11},
\end{equation}

\begin{equation}
Im\bar\Pi=\frac{e^{\beta |k_{0}|}-1}{e^{\beta |k_{0}|}+1}Im\Pi^{11}
\end{equation}

\begin{equation}
Im\bar{\Pi}=-\frac{e^{\beta |k_{0}|}}{2e^{\beta |k_{0}|/2}}Im\Pi^{12}.
\end{equation}
These equations for $ Re\Pi^{11}$ and $ Im\Pi^{11}$ are interesting. It is clear that  
$\Pi^{12}$ and $\Pi^{21}$ are purely imaginary, and
the real part of the gluon self-energy can be obtained from $\Pi^{11}$ or $\Pi^{22}$ [8]. 

  By calculating $\Pi^{12}(k)$, for $\vec{k}\rightarrow 0$, and keeping the energy
$k_{0}\not=0$, we obtain the imaginary part of the $(0,0)$ component 
of the gluon self-energy for the diagram (a) in Fig.1 \\

\begin{eqnarray}
Im\bar{\Pi}(\vec{k}=0, k_{0}) & = & -\frac{e^{\beta |k_{0}|}-1}{2e^{\beta |k_{0}|/2}}Im\Pi^{12}(k_{0}) \nonumber \\
 & = & -\frac{g^{2}Q\delta^{ab}N_{f}}{2\pi}\frac{e^{\beta |k_{0}|}-1}{2e^{\beta |k_{0}|/2}}
\frac{k_{0}^{2}e^{\beta |k_{0}|/2}}{(e^{\beta |k_{0}|/2}+1)^{2}} \nonumber \\
 & = & -\frac{g^{2}Q\delta^{ab}N_{f}}{4\pi}\frac{(e^{\beta |k_{0}|/2}-1)k_{0}^{2}}{(e^{\beta |k_{0}|/2}+1)}
\end{eqnarray}
where we have omitted the $(0,0)$ vector index.

  By calculating $\Pi^{11}(k)$, for $\vec{k}\rightarrow 0$, and keeping the energy $k_{0}\not=0$,
we obtain the real part of the $(0,0)$ component in the vector indices of the gluon
self-energy for the diagram (a) in Fig.1\\

\begin{eqnarray}
Re\bar\Pi(\vec{k}=0,k_{0}) & = & Re\Pi^{11}(k_{0}) \nonumber \\
& = & -\frac{8g^{2}Q\delta^{ab}N_{f}}{\pi^{2}}\int_{0}^{\infty}\frac{\omega^{3}d\omega}
{(e^{\beta\omega}+1)(k_{0}^{2}-4\omega^{2})} 
\end{eqnarray}
where we omitted also the $(0,0)$ vector index.
From Eqs. (31) and (32), we get\\

\begin{equation}
Re\bar{\Pi}(k_{0})=\frac{1}{\pi}\int_{-\infty}^{\infty}\frac{Im\bar{\Pi}(k'_{0})}{k'_{0}-k_{0}}
\mbox{sinh}\theta(\frac{k'_{0}}{2})dk'_{0}
\end{equation}
This is the dispersion relation of the gluon self-energy in Fig.1(a).

\vspace{1cm}

\begin{center}
\section{QUARK POTENTIAL}
\end{center}

  In this section we study the potential between a static quark and 
anti-quark within TFD.\\

  We consider the free energy of a configuration of $N_{q}$ quarks and $N_{\bar{q}}$ 
anti-quarks, $F({\vec{r}}_{1},{\vec{r}}_{2},\dots {\vec{r}}_{N_{q}}; 
{\vec{r}}_{1}^{'},{\vec{r}}_{2}^{'}\dots {{\vec{r}}_{N_{\bar q}}}^{'})$.
By definition\\

\begin{equation}
e^{-\beta F({\vec{r}}_{1},{\vec{r}}_{2},...{\vec{r}}_{N_{q}}; 
{\vec{r}}_{1}^{'},{\vec{r}}_{2}^{'},...{\vec{r}}_{N_{\bar q}}^{'})}
=\frac{1}{N^{(N_{q}+N_{\bar q})}}\sum_{|s\rangle}\langle S|e^{-\beta H}|S\rangle
\end{equation}
where $H$ is Hamiltonian of the system and in the state, $|S\rangle$, there are 
$N_{q}$ static heavy quarks at ${\vec{r}}_{1},...{\vec{r}}_{N_{q}}$ and $N_{\bar q}$ 
static heavy anti-quarks at ${\vec{r}}_{1}^{'},...{\vec{r}}_{N_{\bar q}}^{'}$, and $\sum_{|s\rangle}\langle S|S\rangle =N$.

  We can write\\
  
\begin{eqnarray}
\sum_{|s\rangle}\langle S|e^{-\beta H}|S\rangle & = & \sum_{|s'\rangle}\langle S'|\sum_{(a,b)}
\psi_{a_{1}}({\vec{r}}_{1})\psi_{a_{2}}({\vec{r}}_{2})...\psi_{a_{N_{q}}}({\vec{r}}_{N_{q}})
\psi_{b_{1}}^{c}({\vec{r}}_{1}^{'})\psi_{b_{2}}^{c}({\vec{r}}_{2}^{'})...
\psi_{b_{N_{\bar q}}}^{c}({\vec{r}}_{N_{\bar q}}^{'})
e^{-\beta H} \nonumber \\
& & \psi_{a_{1}}^{\dagger}({\vec{r}}_{1})\psi_{a_{2}}^{\dagger}({\vec{r}}_{2})...
\psi_{a_{N_{q}}}^{\dagger}({\vec{r}}_{N_{q}})\psi_{b_{1}}^{{c}^{\dagger}}({\vec{r}}_{1}^{'})
\psi_{b_{2}}^{{c}^{\dagger}}({\vec{r}}_{2}^{'})...\psi_{b_{N_{\bar q}}}^{{c}^{\dagger}}
({\vec{r}}_{N_{\bar q}}^{'})|S'\rangle
\end{eqnarray}
where these quarks and anti-quarks are taken from $|S\rangle$, and in $|S'\rangle $ there
are no quarks and anti-quarks, and $\psi_{a_{1}}({\vec{r}}_{1})$ $(\psi_{a_{1}}^{\dagger}({\vec{r}}_{1}))$
and $\psi_{b_{1}}^{c}({\vec{r}}_{1}^{'})$ $(\psi_{b_{1}}^{{c}^{\dagger}}({\vec{r}}_{1}^{'}))$
are annihilation ( creation ) operators of quark and anti-quark respectively.  

  From Eqs. (34) and (35) we get \\

\begin{equation}             
e^{-\beta F_{N_{q},N_{\bar q}}}=Tr\lbrack e^{-\beta H}L({\vec{r}}_{1})...L({\vec{r}}_{N_{q}})
L^{\dagger}({\vec{r}}_{1}^{'})...L^{\dagger}({\vec{r}}_{N_{\bar q}}^{'})\rbrack
\end{equation}
where the trace, $Tr$, is $\sum_{|s'\rangle}$, only over states of the pure gauge
field and includes no quark field. The quarks are considered as external particles, 
and the Wilson loop, $L(\vec{r})$, is \\

\begin{equation}
L(\vec{r})=\frac{1}{N}Tr\lbrack\mbox{T}e^{\int_{0}^{t}gT^{a}A_{0}^{a}(\vec{r},\tau)d\tau}\rbrack  
\end{equation}
where $\mbox{T}$ represents time ordering. \\

  For the quark operator, $\psi$, we have \\
  
\begin{equation}
\frac{1}{i}\frac{\partial \psi}{\partial t}=[ H, \psi ]_{-}
\end{equation}

\begin{eqnarray}
H & = & \int d^{3}\vec{x}\cal{H}, \nonumber \\
\cal{H} & = & -ig \bar{\psi} \gamma_{\mu} \vec{\tau} \psi {\vec{A}}_{\mu}
+ m \bar{\psi} \psi=-ig \bar{\psi} \gamma_{0} \vec{\tau} \psi {\vec{A}}_{0}
+ m \bar{\psi} \psi
\end{eqnarray}
due to the quark being static, with the current, $\vec{J}=0$ and $J_{0}\not=0$. 

  Using the commutation relations \\

\begin{equation}
\lbrack\psi_{a}^{\dagger}(\vec{r_{i}},t), \psi_{b}(\vec{r_{j}},t)\rbrack_{+}=\delta_{ij}\delta_{ab},
\end{equation}
we have \\

\begin{equation}
[ H, \psi ]_{-}=-ig \vec{\tau} \vec{A}_{0} \psi + m \psi . 
\end{equation}
We put $m=0$, hence we get \\

\begin{equation}
\frac{1}{i}\frac{\partial\psi(x)}{\partial t}=\tau^{\alpha}A_{0}^{\alpha}(x)\psi(x).
\end{equation}
From Eq. (42), we get \\

\begin{equation}
\psi_{a}(\vec{r},t)=\mbox{T}\hspace{0.1in}exp[\int_{0}^{t}gT^{a}A_{0}^{a}(\vec{r},\tau)d\tau]\psi_{a}(\vec{r},0)
\end{equation}
Then the relation given in Eq. (36) can be derived from Eq. (43). For one static quark, Eq. (36) becomes \\

\begin{equation}
e^{-\beta F_{1,0}}=\mbox{Tr}e^{-\beta H}L(\vec{r})
\end{equation}
and

\begin{equation}
e^{-\beta F_{0,0}}=\mbox{Tr}e^{-\beta H}.
\end{equation}

Hence we have \\

\begin{equation}
e^{-\beta(F_{1,0}-F_{0,0})}=\frac{Tr \lbrack e^{-\beta H}L(\vec{r})\rbrack}
{Tr e^{-\beta H}},
\end{equation}
where $F_{0,0}$ is the free energy of the pure gluon field and $F_{1,0}$ is the free
energy of one quark and the gluon field. Since \\

\begin{equation}
\mbox{Tr}\lbrack e^{-\beta H}L(\vec{r})\rbrack
=\sum_{|s'\rangle}\langle s'| e^{-\beta H} L(\vec{r})|s'\rangle,
\end{equation}

\begin{equation}
\mbox{Tr} e^{-\beta H}=\sum_{|s'\rangle}\langle s'| e^{-\beta H} |s'\rangle
\end{equation}
where the $|s'\rangle$ is the state with only gluons and \\

\begin{equation}
H |s'\rangle = E_{s'} |s^{'}\rangle ,
\end{equation}
we have \\

\begin{equation}
\frac{Tr e^{-\beta H}L(\vec{r})}{Tr e^{-\beta H}}
=\frac{\sum_{|s'\rangle}e^{-\beta E_{s'}}\langle s'|L(\vec{r})|s'\rangle}
{\sum_{|s'\rangle}\langle s'|s'\rangle e^{-\beta E_{s'}}}.
\end{equation}

We can scale $E_{s'}$ up $\alpha$ times and continue $\alpha\beta$ to the complex 
plane, and given that \\

\begin{equation}
\alpha=\rho e^{i\theta}, \hspace{0.2in} \theta=-\epsilon,
\end{equation}
and put $ Re\alpha\rightarrow\infty $  and $ Im\alpha < 0 $, 
then $-\alpha\beta\rightarrow -\infty$, therefore we get \\

\begin{equation}
e^{-\beta F_{q}}=\langle L(\vec{r})\rangle
\end{equation}
where $F_{q}$ is the free energy of a single-quark system without the quanta 
of pure gauge field.

  For a static quark and anti-quark we have \\
  
\begin{equation}
e^{-\beta V(\vec{r}-\vec{r'})} = \langle L^{\dagger}(\vec{r}) L(\vec{r'}) \rangle_{c}
\end{equation}
where $V(\vec{r}-{\vec{r}}^{'})$ is the potential between a static quark and 
anti-quark, and the lower index $c$ representing the connected two-point function of 
the Wilson loop. 

  First we consider the case of the abelian gauge field. This is easily done
in QED, by including the contributions of all Feynman diagrams of virtual photons
exchanged between an electron and a positron. Now we consider the $e\bar{e}$ 
potential energy at finite temperature. In this case, starting with a skeleton 
expansion, in which only full photon propagator and the renormalized 
coupling constant $e$ are used, then the screening potential is the summation
of a set of ladder graphs shown in Fig.2.\\
Therefore we get \\

\begin{equation}
e^{-\beta V(r,\beta)} = 1 + \sum_{n=1}^{\infty}\frac{1}{n!}\lbrack\frac{e^{2}\beta}{4\pi r}
e^{-m(\beta)r}\rbrack^{n}\\
=exp\lbrack\frac{e^{2}\beta}{4\pi r}e^{-m(\beta)r}\rbrack.
\end{equation}
This simply leads to a screened Coulomb potential \\

\begin{equation}
V(r,\beta)=-\frac{e^{2}}{4\pi r}e^{-m(\beta)r},
\end{equation}
where $m(\beta)$ is the renormalized screening mass of a photon at finite temperature. 
The magnitude of $m(\beta)$ is given by an expression similar to Eq. (16). \\
  Now we study the case of QCD in TFD. For the connected two-point function of Wilson
loop in Eq. (53), up to $O(g^{4})$ the contributions are from the diagrams in Fig.3.  

  The contribution of the one-gluon exchanged diagram $3(a)$ to the connected
two-point function of Wilson loop equals zero since \\

\begin{equation}
\mbox{Tr}(T^{a})=0.
\end{equation}
For the two diagrams $3(b)$ and $3(c)$, using the thermal gluon propagator
in Eqs. (4) and (5), we have obtained the contribution of $O(g^{4})$ 
to the connected two-point function of the 
thermal Wilson loop as follows: \\

\begin{equation}
\mbox{G}(\vec{p})=\frac{g^{4}Q^{2}(N^{2}-1)}{2^{7}\pi^{3}T^{2}}\frac{1}{|\vec{p}|},
\end{equation}
where $\mbox{Tr}(T^{a}T^{b})=Q\delta^{ab}$, $g$ is the coupling constant, $T$
is the temperature, and the factor $(N^{2}-1)$ is from the adjoint representation 
of the gluon for the gauge group $SU(N)$. Here we have used the following formulas:\\

\begin{equation}
\int_{0}^{\infty}\frac{[(1+a)e^{x}+a]}{(1+e^{x})^{2}}e^{-ax}x^{n}dx=n!\sum_{k=1}^{\infty}
\frac{(-1)^{k+1}}{(a+k)^{n}},
\end{equation}
which has the value \\

\begin{equation}
\sum_{k=1}^{\infty}\frac{(-1)^{k+1}}{k^{2}}=\frac{\pi^{2}}{12};
\end{equation} 
for $a=0$ and $n=2$ \\
and \\

\begin{equation}
\int\frac{d^{n}k}{(2\pi)^{n}}\frac{1}{(k^{2})^{\alpha}[(p-k)^{2}]^{\beta}}
=\frac{1}{(16\pi^{2})^{\frac{n}{4}}}(p^{2})^{-\alpha-\beta+\frac{n}{2}}
\frac{\Gamma(\alpha+\beta-\frac{n}{2})}{\Gamma(\alpha)\Gamma(\beta)}
B(\frac{n}{2}-\beta,\frac{n}{2}-\alpha),
\end{equation}
where $B(\frac{n}{2}-\beta,\frac{n}{2}-\alpha)$ is the Beta function.

  The function $\mbox{G}(\vec{p})$, in Eq. (57), can be 
Fourier transformed to the coordinate space, as follows: \\

\begin{equation}
\mbox{G}(\vec{x}-\vec{x'})=\frac{g^{4}Q^{2}(N^{2}-1)}{2^{5}\pi^{2}T^{2}}\frac{1}{(\vec{x}-\vec{x'})^{2}}.
\end{equation} \\

  If we take into account the thermal gluons in Fig.3 with the color-electric
pole mass, $m_{el}(T)$ and use the following formula \\

\begin{equation}
\int\frac{d^{3}\vec{k}d^{3}\vec{p}\mbox{e}^{-i\vec{p}(\vec{x}- {\vec{x}}^{'})}}
{({\vec{k}}^{2}+m_{el}^{2}(r))[(\vec{p}-\vec{k})^{2}+m_{el}^{2}(r)]}
=\frac{4\pi^{4}}{(\vec{x}- {\vec{x}}^{'})^{2}}
\mbox{e}^{-2m_{el}(T)|\vec{x}- {\vec{x}}^{'}|},
\end{equation}
we get \\

\begin{equation}
\mbox{G}(\vec{x}- {\vec{x}}^{'})=\frac{g^{4}Q^{2}(N^{2}-1)}{2^{5}\pi^{2}T^{2}}
\frac{\mbox{e}^{-2m_{el}(T)|\vec{x}- {\vec{x}}^{'}|}}{(\vec{x}- {\vec{x}}^{'})^{2}}.
\end{equation}
Therefore the potential between a quark and an anti-quark at high 
temperature up to $O(g^{4})$ is \\

\begin{equation}
\mbox{V}(r)=\frac{-g^{4}Q^{2}(N^{2}-1)}{2^{5}\pi^{2}T}
\frac{\mbox{e}^{-2m_{el}(T)r}}{r^{2}}.
\end{equation}

  It is clear from Eqs. (55) and (64) that the potential in a
QCD plasma is different than the potential in a QED plasma, though they
all decrease exponentially with the increase of the distance, $r$, between
them, but the decrease in power are different, the latter is linear while the
former is quadratic [9].

\vspace{1cm}

\section{CONCLUSIONS AND DISCUSSIOS}

  In summary, we have obtained several consequences for QGP within TFD framework:    
the screening length in QGP in one-loop order, the dispersion relations between 
the real part and the imaginary part of the thermal gluon self-energy and the 
quark potential at high temperature up to $O(g^{4})$. The main advantage of TFD 
compared to other formalisms of thermal field theories is the ease of calculation. 

  Hopefully QGP will be observed by experiments at CERN and at BNL-RHIC in the
near future. As an estimate, the binding radius $R_{B}$ of $J/\psi$ particle [10] 
can be considered to be roughly $R_{B}\approx 0.3 fm$ [11], setting the running coupling 
constant $g(T)\simeq 1.0$, then, we obtain the solution $L(T)=R_{B}$ and the 
critical temperature $T_{c}\approx 0.3 Gev$. This means that a suppression of 
$J/\psi$ mesons should be observed if the temperature in a nuclear collision 
becomes much higher than the transition temperature $T_{c}$. \\

  This research is supported in part by Natural Sciences and Engineering
  Research Council of Canada.

\end{document}